\documentclass[twoside,twocolumn]{article}

\usepackage{blindtext} 

\usepackage[sc]{mathpazo} 
\usepackage[T1]{fontenc} 
\usepackage{microtype} 

\usepackage[english]{babel} 

\usepackage[hmarginratio=1:1,top=2cm, bottom=2cm, left=2cm, right=2cm,columnsep=20pt]{geometry} 
\usepackage[hang, small,labelfont=bf,up,textfont=it,up]{caption} 
\usepackage{booktabs} 

\usepackage{lettrine} 

\usepackage{enumitem} 
\setlist[itemize]{noitemsep} 

\usepackage{abstract} 

\usepackage{titlesec} 
\titleformat{\section}[block]{\large\scshape}{\thesection.}{1em}{} 
\titleformat{\subsection}[block]{\large}{\thesubsection.}{1em}{} 

\usepackage{fancyhdr} 
\usepackage{titling} 

\usepackage{hyperref} 
\usepackage{mathtools}
\usepackage{tikz}
\usepackage{verbatim}
\usepackage{mathrsfs}
\usepackage{gensymb}
\usepackage{subfigure}
\usepackage{makecell}

\pretitle{\begin{center}\Huge\bfseries} 
\posttitle{\end{center}} 
\title{Fast Real-Time Shading for Polygonal Hair} 
\author{%
\textsc{Martin Gerard} \\[1ex] 
\normalsize Eko Software \\ 
}
\date{} 
\renewcommand{\maketitlehookd}{%
\begin{abstract}
Though a lot of improvement has been made to hair rendering techniques in the recent years, realistic rendering of hair remains a challenge, especially in real time. In this paper, we propose a fast technique to approximate the shading of hair lighted by an environment map, direct lighting or a global illumination system, without having to render deep opacity maps or requiring additional artistic work. 
\begin{center}
\includegraphics[scale = 1]{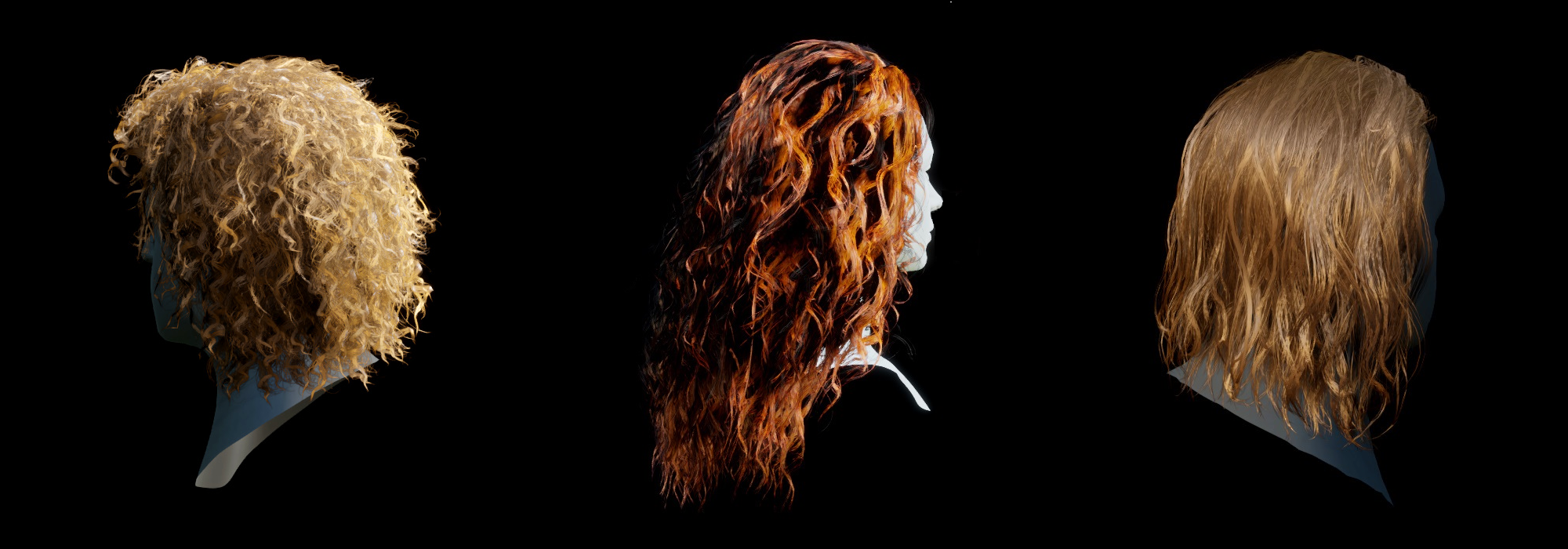}
\end{center}
\end{abstract}
}

\begin{document}

\maketitle


\section{Introduction}

Hair is one of the most difficult material to render in real-time, yet an extremely important visual component of characters and can easily break immersion if not done properly. This has lead for artists and programmers to come up with various solutions that in some cases require careful tweaking in order to achieve the desired realism in the dedicated time budget, and add a significant amount of time-consuming work for both artists and programmers. Most of these techniques require to maintain data structures at run-time, like deep opacity maps, to represent volumetric self-shadowing and can be costly.

\hfill

In this paper, we present a novel method for shading hair realistically with dynamic lights or a GI system in a single pass, without having to maintain a dedicated data structure, and without the need of additional textures other than a diffuse map and a flow map. 

\hfill

Our work builds upon the model of a phase function of a single hair by \cite{Marschner:2003dg} through the addition of a self-shadowing term baked in the geometry and a technique for a fast integration of the model over an environment map.


\section{The hair rendering equation}

Let's begin by writing the rendering equation for a hair of tangent $u$

\begin{equation}
\begin{multlined}
L_{out}(x, \omega_o) = 
\int_{\Omega}{L_{in}(x, \omega_i)S(\omega_i, \omega_o, u)d\omega_i}
\end{multlined}
\end{equation}

$L_{out}(x, \omega_o)$ and $L_{in}(x, \omega_i)$ are respectively the radiance going out of point $x$ in direction $\omega_o$ and the radiance coming to x from direction $\omega_i$. $S(\omega_o, \omega_i, u(x))$ is the single-hair phase function expressing the fraction of the light coming from direction $\omega_i$ that gets scattered in direction $\omega_o$ by a hair of tangent $u$.

\hfill

We approximate the input radiance field $L_{in}(x, \omega_i)$ by the product of a far field $L_{far}(\omega_i)$ with a near field transmittance function $V(x, \omega_i)$:

\begin{equation}
L_{in}(x, \omega_i) = L_{far}(x, \omega_i)V(x, \omega_i)
\end{equation}

The rendering equation becomes:

\begin{equation}
\begin{multlined}
L_{out}(x, \omega_o) = \int_\Omega{L_{far}(x, \omega_i)V(x, \omega_i)S(\omega_i, \omega_o, u)d\omega_i}
\end{multlined}
\end{equation}

\subsection{The single hair phase function}

Let's focus first on the single-hair phase function $S(\omega_i, \omega_o, u)$

\begin{figure}[h]
	\centering
	\includegraphics[scale = 0.18]{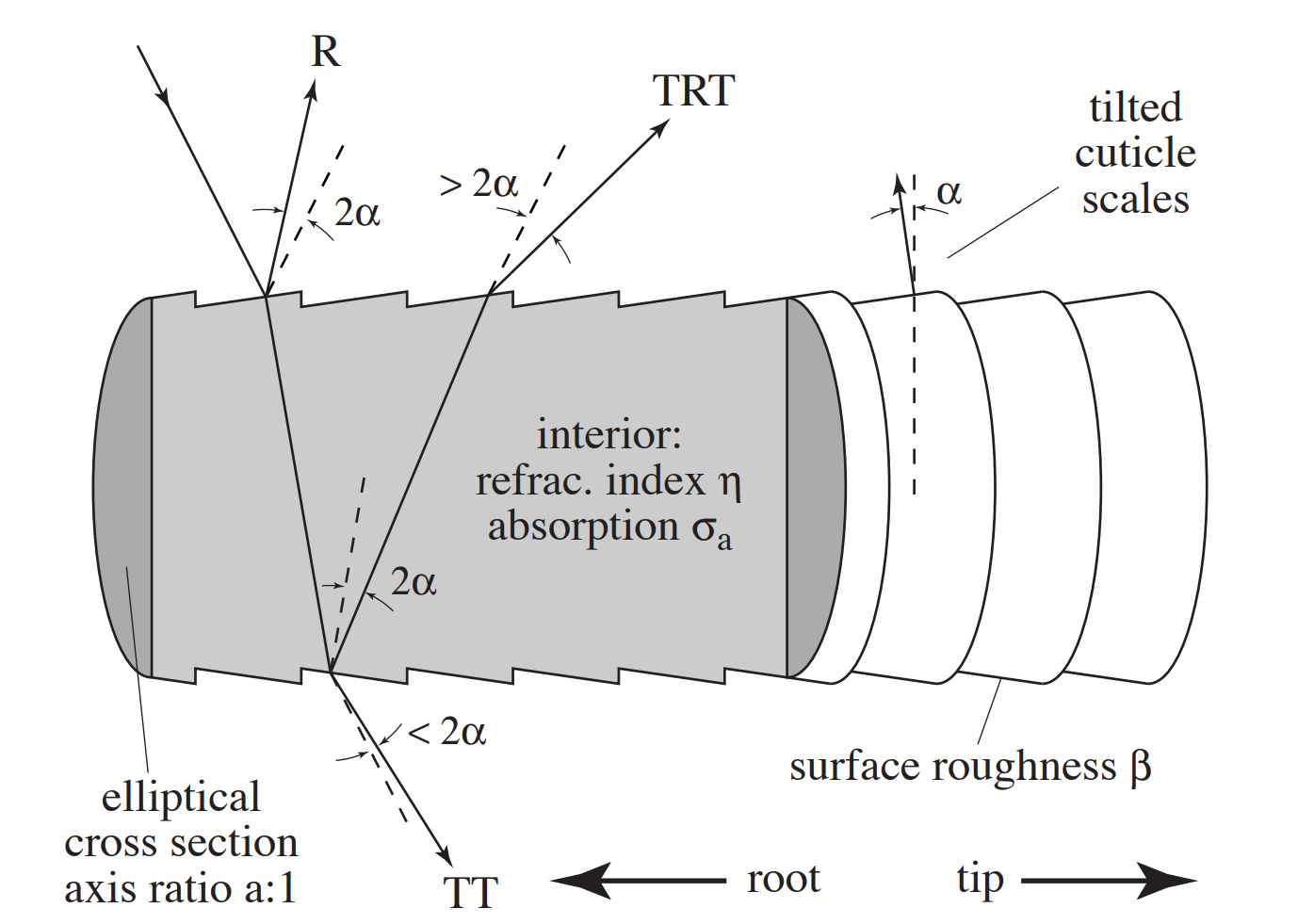}
	\caption{A schematic of our model for a hair fiber. The dashed lines indicate the scattering angles for a cylinder without tilted surface scales.}
	\label{fig:fig1}
\end{figure}

In \cite{Marschner:2003dg}, the authors separate the phase function in three terms, corresponding to the light directly reflected on the surface of the hair (the R mode), the light transmitted into the hair then transmitted out of the hair through the other side (the TT mode), and the light transmitted into the hair, reflected on the inner surface of the hair then transmitted outside of the hair (the TRT mode) as represented in figure \ref{fig:fig1}.

\begin{figure}[h]
	\centering
		\includegraphics[scale=0.18]{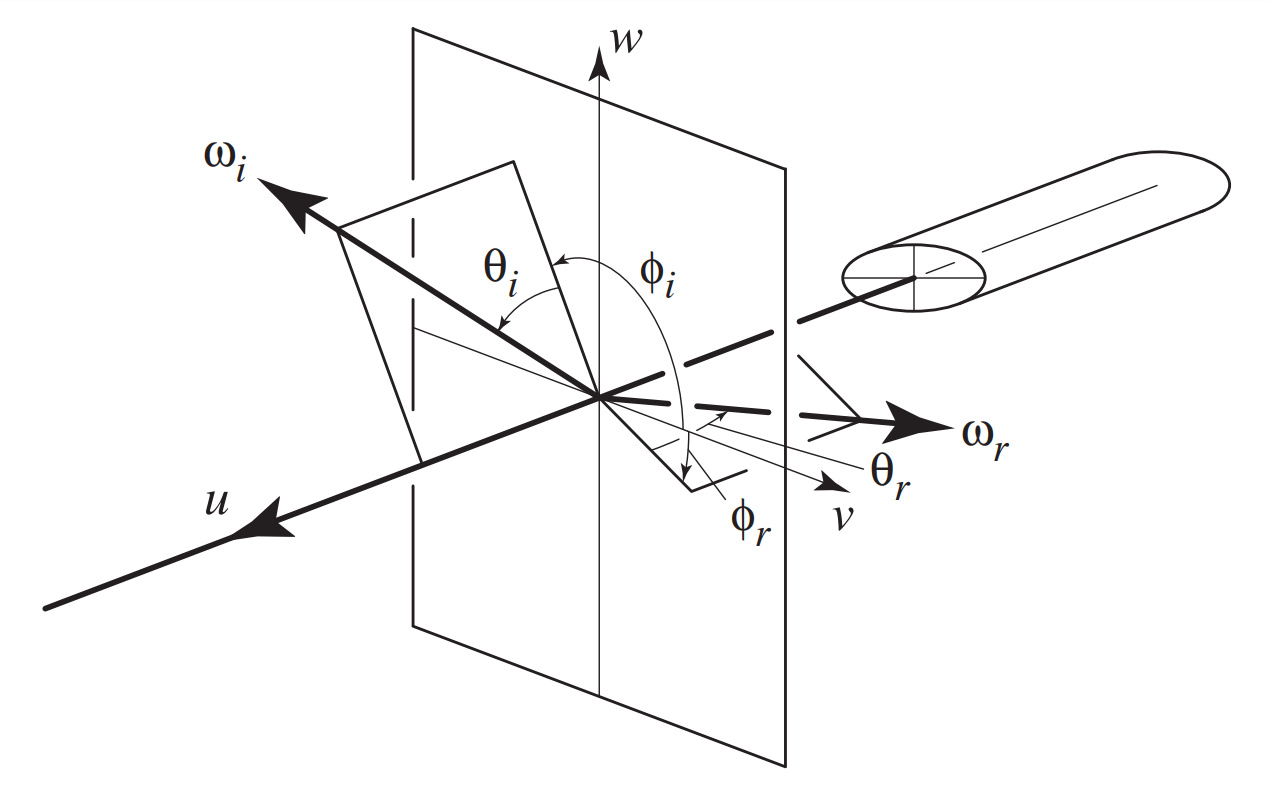}
		\caption{Notation for scattering geometry}
	\label{fig:fig2}
\end{figure}

We denote by $\phi_i$ and $\phi_r$ the azimuthal angle of the incident ray and the outgoing ray in the normal plane of the hair. $\theta_i$ and $\theta_r$ refer to the inclination of the same rays with respect to the normal plane. We will also define the difference angle $\theta_d = (\theta_r - \theta_i)/2$, the half angles $\theta_h = (\theta_i + \theta_r)/2$ and $\phi_h = (\phi_i + \phi_r)/2$ and the relative azimuth $\phi = \phi_r - \phi_i$ . The angles are drawn in figure \ref{fig:fig2}.

\hfill

In these conditions, \cite{Marschner:2003dg} shows that the phase function $S(\omega_i, \omega_r, u)$ can be expressed as

\begin{equation}
\begin{aligned}
S(\omega_i, \omega_r, u) &= \\
&M_R(\theta_h)N_R(\eta, \theta_d, \phi)/cos^2\theta_d + \\
&M_{TT}(\theta_h)N_{TT}(\theta_d, \phi)/cos^2\theta_d + \\
&M_{TRT}(\theta_h)N_{TRT}(\theta_d, \phi)/cos^2\theta_d
\end{aligned}
\end{equation}

\hfill

where $M_R$, $M_{TT}$ and $M_{TRT}$ are the three following longitudinal functions

\begin{equation}
\begin{aligned}
M_R(\theta_h) &= \mathcal{G}(\beta_R, \theta_h - \alpha_R)\\
M_{TT}(\theta_h) &= \mathcal{G}(\beta_{TT}, \theta_h - \alpha_{TT})\\
M_{TRT}(\theta_h) &= \mathcal{G}(\beta_{TRT}, \theta_h - \alpha_{TRT})\\
\end{aligned}
\end{equation}

\hfill

where $\mathcal{G}(\beta, x)$ is a unit-integral, zero-mean lobe function of width $\beta$. We chose to use a Henyey-Greenstein function width $\beta$, which is faster to compute than a Gaussian.

\begin{equation}
\mathcal{G}(\beta, cos\theta_h) = \frac{1}{4\pi} \frac{1 - g(\beta)^2}{(1 + g(\beta)^2 - 2g(\beta)cos\theta_h)^{\frac{3}{2}}}
\end{equation}

\cite{Marschner:2003dg} gives the following surface values

\begin{equation*}
\begin{aligned}[c]
\beta_R &= 5\degree\textrm{ to }10\degree\\
\beta_{TT} &= \beta_R / 2\\
\beta_{TRT} &= 2\beta_R
\end{aligned}
\textrm{  and  }
\begin{aligned}[c]
\alpha_R &= -5\degree\textrm{ to }-10\degree\\
\alpha_{TT} &= -\alpha_R/2\\
\alpha_{TRT} &= -3\alpha_R/2\\
\end{aligned}
\end{equation*}

For $\beta_R = 10\degree$ and $\alpha_R = -10\degree$, we find the following values for $g(\beta)$ by minimizing the $L_2$ norm of the difference between $\mathcal{G}$ and a gaussian.

\begin{align*}
g(\beta_R) &= 0.752\\
g(\beta_{TT}) &= 0.865\\
g(\beta_{TRT}) &= 0.578\\
\end{align*}

The azimuthal part $N(\theta_d, \phi)$ is detailed in \cite{Marschner:2003dg} and is to costly to evaluate analytically. We use the following approximation

\begin{equation}
\begin{aligned}
N_p(\theta_d, \phi) &= N_{p,LUT}(\theta_d, \phi)T(\sigma_a, h_{max})^p
\end{aligned}
\end{equation}

where $N_{p,LUT}$ is a look-up texture generated using the algorithms described by \cite{Marschner:2003dg}, $\sigma_a$ is the colored absorption of the hair and $h_{max}$ is the offset of the incident light from the fiber that maximizes $T(\sigma_a, h)$. We get

\begin{equation}
\begin{aligned}
\label{eqn:azimuth}
N_R(\theta_d, \phi) &= N_{R,LUT}(\theta_d, \phi)\\
N_{TT}(\theta_d, \phi) &= N_{TT,LUT}(\theta_d, \phi)e^{-4\sigma_a}\\
N_{TRT}(\theta_d, \phi) &= N_{TRT,LUT}(\theta_d, \phi)e^{-8\sigma_a(1 - \frac{3}{2\eta^2})}\\\\
\textrm{where}\\
\eta &= \sqrt{\frac{1.55^2 - cos^2\theta_d + 1}{cos\theta_d}}
\end{aligned}
\end{equation}


\subsection{Per-vertex near field transmittance}

We now have a mean to evaluate the phase function for a single hair, but it doesn't take into account the presence of surrounding hair. For simplicity, we don't take multiple scattering into account, and simply focus on the occlusion of the light from the surrounding hair. This allows us to consider a function $V(\omega_i)$ that only depends on the geometry itself and thus can be precomputed for static hair. One could object that hair are rarely static but we find that the precomputed data still give plausible results for small deformations.

\hfill

We precompute the near field transmittance by rendering a cubemap for each vertex that stores the transmittance for each direction using alpha blending. A transmittance of 1 means that the vertex is unoccluded in that direction whereas a transmittance of 0 means that it is entirely shadowed. We then store a spherical harmonics approximation of the cubemap in the vertex attributes.

\hfill

At runtime, we extract the spherical harmonics coefficients in the vertex shader and rotate them to take into account the hair orientation before passing them to the pixel shader.

\hfill

This technique allows us to evaluate the obstruction of light by the surrounding hair at any point in any given direction simply by decompressing the vertex attributes, without having to render deep opacity maps for every light.

\subsection{Direct lighting}

For a point light or a directional light, the lighting equation simplifies to 

\begin{equation}
L_{direct}(\omega_o, x) = L_{light}(x)V(\omega_l, x)S(\omega_o, \omega_l, u)
\end{equation}

where $L_{light}(x)$ is the radiance that would be received from the light at point $x$ if there were no hair, and $\omega_l$ is the direction from the point $x$ to the light. $V(\omega_l, x)$ can be evaluated directly from its spherical harmonics decomposition.

\hfill

With a low order approximation of $V$, the overestimation of $V$ applied to the transmission mode can produce undesirable light leaking from back lights, in that case, this issue can be alleviated by introducing a bias such that 

\begin{equation}
\tilde{V}(\omega_l, x) = max\left(0, \frac{V(\omega_l, x) - bias}{1 - bias}\right)
\end{equation}


\subsection{Integration over the far field}

In what follows, we will assume that we have access to a spherical harmonics representation of the far field. Given that the transmittance function $V$ is already expressed as spherical harmonics, we propose to decompose the phase function in spherical harmonics so that the rendering equation can be reduced to the integral of the product of three spherical harmonics, which can thus be computed as a combination of their coefficients.

\begin{equation}
S_{\omega_r}(\omega_i) = S^R_{\omega_r}(\omega_i) + S^{TT}_{\omega_r}(\omega_i, \sigma_a) + S^{TRT}_{\omega_r}(\omega_i, \sigma_a)
\end{equation}

It follows from equation \ref{eqn:azimuth} that 

\begin{equation}
S^{TT}_{\omega_r}(\omega_i, \sigma_a) = S^{TT}_{\omega_r}(\omega_i)e^{-4\sigma_a}
\end{equation}

We find that the $S^{TRT}_{\omega_r}$ can be approximated by

\begin{equation}
S^{TRT}_{\omega_r}(\omega_i, \sigma_a) = S^{TRT}_{\omega_r}(\omega_i)e^{-\frac{11}{2}\sigma_a}
\end{equation}

So that

\begin{equation}
S_{\omega_r}(\omega_i) = S^R_{\omega_r}(\omega_i) + S^{TT}_{\omega_r}(\omega_i)e^{-4\sigma_a} + S^{TRT}_{\omega_r}(\omega_i)e^{-\frac{11}{2}\sigma_a} 
\end{equation}

Let $a^R_{\omega_r}$, $a^{TT}_{\omega_r}$ and $a^{TRT}_{\omega_r}$ be the coefficients of the spherical harmonics decomposition of $S^R_{\omega_r}$, $S^{TT}_{\omega_r}$ and $S^{TRT}_{\omega_r}$ respectively, then 

\begin{equation}
S_{\omega_r}(\omega_i) = \sum \left[ a^R_{\omega_r} + a^{TT}_{\omega_r} e^{-4\sigma_a} + a^{TRT}_{\omega_r}e^{-\frac{11}{2}\sigma_a}  \right] Y_{lm}(\omega_i)
\end{equation}

\begin{figure*}
\centering 
\label{fig:render}
\begin{center}
\begin{tabular}{|*{4}{m{0.18\textwidth}}|}
\hline
\centering & \makecell{Kajiya-Kay \\ + Ambient term} & \makecell{Ours} & \makecell{Ground Truth} \\
\makecell{Environment map} & \makecell{\includegraphics[scale=0.2]{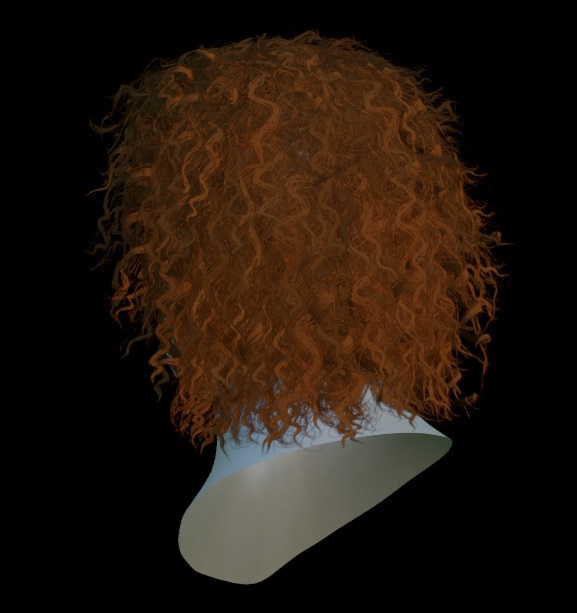}} & \makecell{\includegraphics[scale=0.2]{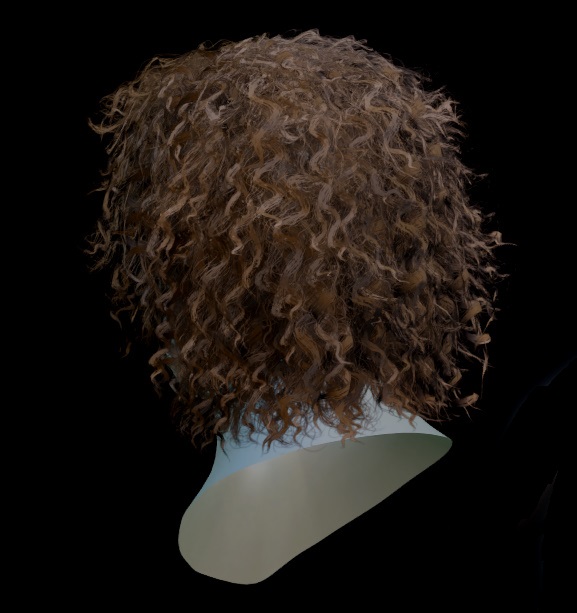}} & \makecell{\includegraphics[scale=0.2]{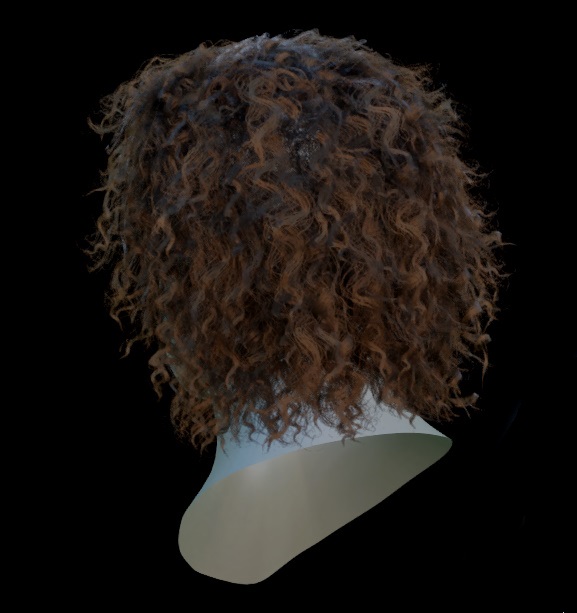}} \\
\makecell{Environment map\\ + Directional light} & \makecell{\includegraphics[scale=0.2]{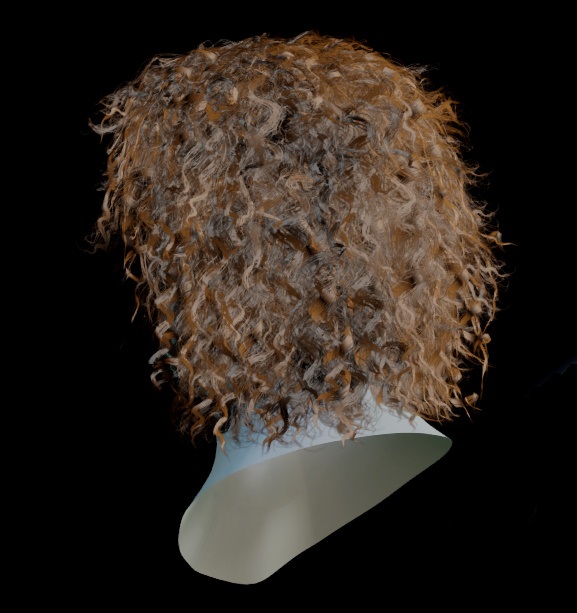}} & \makecell{\includegraphics[scale=0.2]{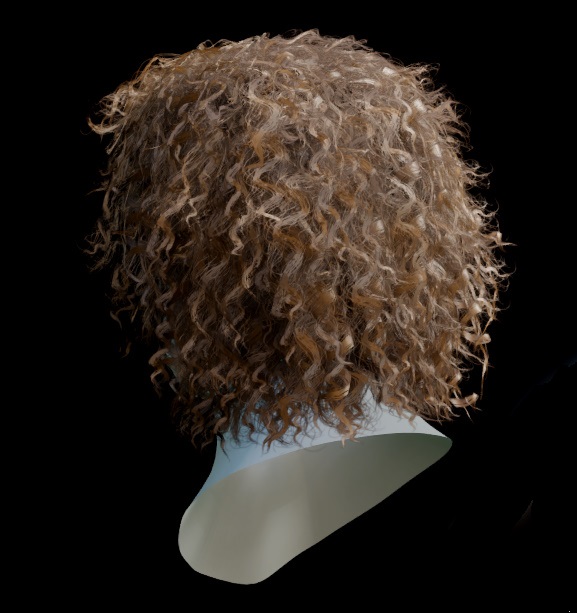}} & \makecell{\includegraphics[scale=0.2]{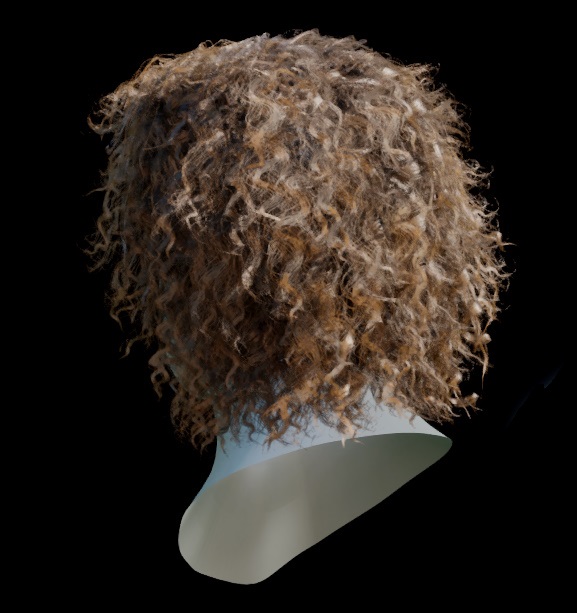}}\\
\hline
\end{tabular}
\caption{Polygonal hair rendered with our technique versus Kajiya-Kay in a comparable amount of time. We compare it to a path-traced reference render using the single scattering approximation.}
\end{center}
\end{figure*}

\hfill

$a^R$, $a^{TT}$ and $a^{TRT}$ are scalar functions of $\omega_r$ and can be precomputed. By choosing to express the coefficients in a base where the hair tangent $u$ is pointing up, we can store each coefficient in a 1D look-up texture parametrized by $sin(\theta_r)$.

\hfill

At runtime, we can sample the look-up texture, rotate the coefficients back to world space, then evaluate the triple product integral as a sum of integrals of spherical harmonics that can be computed using Wigner 3-j symbols in the following way

\begin{equation}
\begin{split}
&\int_{\Omega} Y_{l_1m_1}(\omega)Y_{l_2m_2}(\omega)Y_{l_3m_3}(\omega) d\omega = \\
&\sqrt{\frac{(2l_1 + 1)(2l_2 + 1)(2l_3 + 1)}{4\pi}}
\begin{pmatrix}
l_1 & l_2 & l_3 \\
0 & 0 & 0 \\
\end{pmatrix}
\begin{pmatrix}
l_1 & l_2 & l_3 \\
m_1 & m_2 & m_3 \\
\end{pmatrix}
\end{split}
\end{equation}

\section{Implementation}

We used 2-band SH (9 scalar coefficients) for $V$ encoded in the vertex attributes and 1-band SH for the phase function. The phase function look-up texture is a 128x3 8bits normalized RGBA texture, parametrized by $(sin(\theta_r) + 1)/2$ in x.
For hair transparency, we used Adaptive Transparency by \cite{Salvi:2011dg} with 4 nodes.

\begin{figure}[h]
\centering
\begin{tabular}{|c|c|c|}
\hline
& \makecell{Kayija-Kay} & \makecell{Ours}\\
\hline
\makecell{Curly hair (9K tri) \\ Close up} & 0.138 ms & 0.316 ms\\ 
\hline
\makecell{Long curly hair (34K tri) \\ Close up} & 0.248 ms & 0.458 ms\\
\hline
\makecell{Curly hair (9K tri) \\ Larger shot} & 0.036 ms & 0.068 ms \\
\hline
\makecell{Long curly hair (34K tri) \\ Larger shot} & 0.040 ms & 0.081 ms \\
\hline
\end{tabular}
\caption{Render time of two hair model in close up at 1080p on a NVIDIA RTX 3080 with our technique.}
\end{figure}


\section{Results}

We use a path-traced reference which we call "ground truth" (without taking into account multiple scattering) which we compare against our technique in Fig. 3.

All results are measured on a NVIDIA RTX 3080 at 1080p with two different scene settings. The first setting is a close up shot, where the hair itself covers a surface of approximately 600x600 pixels whereas the second setting is a larger shot, more representative of a common case in a video game where the hair covers a surface of approximately 150x150 pixels. In both case, the hair are lit by an environment map and a directional light. We compare our timings with \cite{Kajiya:1989dg}'s shading in the same conditions. The results are shown in Fig. 4 and show that our technique is not significantly more expensive than \cite{Kajiya:1989dg} (which is broadly used for being fast) while simulating more features, without the overhead cost of having to render deep opacity maps or maintaining another data structure. 

\section{Conclusions}

In this paper we have presented a fast way of solving the hair rending equation at runtime without maintaining a data structure. The visual result, while preserving the main features of the path-traced reference, still exhibits some differences that can be reduced by increasing the order of the spherical harmonics as needed.




\begin{thebibliography}{99} 

\bibitem[Marschner et al.]{Marschner:2003dg}
Stephen R. Marschner, Henrik Wann Jensen and Mike Cammarano (2003).
\newblock Light Scattering from Human Hair Fibers.
\newblock {\em ACM Transactions On Graphics}, 22,3:780-791.

\bibitem[Salvi and al.]{Salvi:2011dg}
Marco Salvi, Jefferson Montgomery and Aaron Lefohn (2011).
\newblock Adaptive Transparency.
\newblock {\em Conference: Proceedings of the ACM SIGGRAPH/EUROGRAPHICS Conference on High Performance Graphics 2011, Vancouver, Canada, August 5-7, 2011}.

\bibitem[J. Kajiya and T. Kay]{Kajiya:1989dg}
J. Kajiya and T. Kay (1989).
\newblock Rendering fur with
three dimensional textures.
\newblock {\em In SIGGRAPH 89 Conference Proceedings, pp. 271-280, 1989.}.
 
\end{thebibliography}
\end{document}